\documentstyle[sprocl]{article}

\input{psfig}

\def\Journal#1#2#3#4{{#1} {\bf #2}, #3 (#4)}

\def\NPB{{\em Nucl. Phys.} B}
\def\PLB{{\em Phys. Lett.}  B}

\def\ZPC{{\em Z. Phys.} C}
\def\IJMF{{\em Int. J. Mod. Phys.} A4}

\def\be{\begin{equation}}
\def\ee{\end{equation}}
\def\bea{\begin{eqnarray}}
\def\eea{\end{eqnarray}}

\begin{document}

\title{ Tests of the JETSET Bose-Einstein correlation model in the
  $e^+e^-$ annihilation process }

\author{\underline{O.Smirnova}, B.L\"orstad, R.Mure\c san}

\address{Department of Physics, Elementary
  Particle Physics,\\ Box 118, 221 00 Lund, Sweden\\E-mail:
  oxana@quark.lu.se}


\maketitle\abstracts{
Studies of implementation of the simulation of the Bose-Einstein
correlation effects in the JETSET particle generator are performed.
Analysis of dependance of the one-dimensional correlation function
parameters on the presumed boson source size reveals appearance of the
effective new length scale, which limits applicability of the simple
momentum-shifting mechanism, employed by JETSET to simulate the
Bose-Einstein correlation. Two- and three-dimensional correlation
functions are analysed as well and compared to the DELPHI data.  
}
  

\section{Introduction}
\label{sec:intro}

Recent interest in profound studies of the Bose-Einstein correlations
(BEC in what follows) was sparked by reports on possible influence of
this phenomenon on the measured value of the $W$ boson mass in
$e^+e^-$ annihilation~\cite{ww1,ww2}. Estimations of the strength of
this influence were done using the Monte Carlo particle generator
JETSET~\cite{jetset}, which includes a phenomenological model for
two-particle BEC simulation. JETSET (together with
PYTHIA~\cite{jetset}) is so far the only $e^+e^-$ annihilation event
generator which accounts for BEC.  It is known to reproduce well
majority of experimental data, including some basic features connected
to BEC. However, more sophisticated studies reveal not only certain
discrepancies between experimental and model results, but also
self-inconsistencies in the model itself~\cite{haywood,fialkowski}.
Therefore, it is of particular importance to establish the extent of
applicability of this model and to have a full understanding of its
advantages and drawbacks.

This work is devoted to studies of the built-in JETSET algorithm which
is used to simulate BEC in hadronic decays of $Z$ boson. The following
section describes the physical process and corresponding observables.
Overview of the studied model is presented in the next section. The
last section contains results and discussion.


\section{Scope of the study and definitions}
\label{sec:scope}

The goal of this analysis is to study methods and consequences of
implementing Bose-Einstein correlation models in the JETSET event
generator, and to find out eventually how can it affect data analysis.
To accomplish this goal, Bose-Einstein correlations between pions
produced in hadronic decays of the $Z$ boson are investigated. These
decays are generated by the JETSET event generator, which uses the
so-called Lund string model to simulate electron-positron annihilation
events at a given center of mass energy, in our
case $E_{cm}=91.2\ GeV$.

 Only the correlations between pairs of particles are studied, whith the
two-particle correlation function is defined as 
\be
\label{eq:cfdef}
C_2(p_1,p_2)=\frac{P(p_1,p_2)}{P(p_1)P(p_2)}\ ,
\ee
where $p_1$ and $p_2$ are four-momenta of two particles, $P(p_1,p_2)$
-- two-particle probability density, and $P(p_1)$ and $P(p_2)$
denotes single-particle probability densities.

Defined as above $C_2$ is parameterized in terms of the invariant four-momenta
difference {\footnotesize $Q=\sqrt{(\vec{p}_1-\vec{p}_2)^2-(E_1-E_2)^2}$} as 
\be
\label{eq:cfpar}
C_2(Q)=N(1+\lambda e^{-R^2Q^2})\ .
\ee
This parameterization is one of the most commonly used~\cite{bengt},
with the parameter $R$ giving the width (or source size) and $\lambda$ -- the
strength of the correlation.

In order to be able to compare our results with experimental data,
only charged pions were used in the analysis. Although prompt pions
produced in the string decay are the most clean sample, we selected only pions
produced in decays of prompt $\rho$ mesons, due to the
following advantages~:\\
\indent -- high sample uniformity: no admixture of particles which are not
  subject to Bose-Einstein correlations;\\
\indent -- sufficient average multiplicity of the sample: more pions are
  produced in decays of $\rho$ mesons then promptly from the string.

Studies were performed not only for the one-dimensional correlation
function $C_2(Q)$, but also for two- and three-dimensional cases. To
facilitate calculations, the Longitudinal Centre-of-Mass
System~\cite{lcms,ox2} (LCMS) was used to measure four-momentum
difference. This is the system in which the sum of the two particles
momenta is perpendicular to the jet axis, hence only two-jet
$q\overline{q}$ events were generated for further simplicity. In LCMS,
$Q$ is resolved into $Q_{long}$, parallel to the jet axis,
$Q_{t,out}$, collinear with the pair momentum sum, and complementary
$Q_{t,side}$, perpendicular to both $Q_{long}$ and $Q_{t,out}$. It is
more convenient in some cases to use only two-dimensional picture with
longitudinal $Q_{\parallel}\equiv Q_{long}$ and perpendicular
$Q_{\perp}=\sqrt{Q_{t,out}^2+Q_{t,side}^2}$. Parameterization
of $C_2$ for these two- and three-dimensional cases was chosen
correspondingly as
\be
\label{eq:multi2}
C_2(Q_{\perp},Q_{\parallel})=N(1+\lambda e^{-Q^2_{\perp}R^2_{\perp}-Q^2_{\parallel}R^2_{\parallel}})\ ,
\ee
\be
\label{eq:multi3}
C_2(Q_{t,out},Q_{t,side},Q_{long})=N(1+\lambda e^{-Q^2_{t,out}R^2_{t,out}-Q^2_{t,side}R^2_{t,side}-Q^2_{long}R^2_{long}})\ .
\ee

In high-energy physics experiments involving detectors, it is
difficult to construct the product $P(p_1)P(p_2)$ from
Eq.(\ref{eq:cfdef}) due to the phase space limitations. Therefore it
is often replaced by $P_0(p_1,p_2)$, which is equal to $P(p_1)P(p_2)$
in a hypothetical case of absence of all the correlations. To make our
results comparable with experiment, we must construct a reference
sample corresponding to $P_0(p_1,p_2)$. Therefore, the measured
two-particle correlation function is calculated as the double-ratio,
using the event mixing technique~\cite{ox2}~: 
\be
\label{eq:double}
r_{BE}(Q)=\frac{N^{\pm\pm}_{BE}(Q)}{N^{\pm\pm}_{BE,mix}(Q)}\ ,\ 
r_{noBE}(Q)=\frac{N^{\pm\pm}_{noBE}(Q)}{N^{\pm\pm}_{noBE,mix}(Q)}\ ,\ 
C_2(Q)=\frac{r_{BE}(Q)}{r_{noBE}(Q)} 
\ee 
Here $N^{\pm\pm}_{BE}(Q)$ is number of like charged pions as a
function of the four-momenta difference $Q$ in presence of
Bose-Einstein correlations. Subscript ``$BE,mix$'' denotes same quantity but
with pairs of pions picked from different events. Indices ``$noBE$'' and
``$noBE,mix$'' correspond to analogous quantities in absence of BEC
(i.e., the simulation of BEC is not included into the event
generation).
 

\section{Model description}
\label{sec:model}

As it was already mentioned, JETSET is the only particle generator
which allows and actually includes an algorithm emulating
Bose-Einstein correlations. Recall that BEC is the quantum mechanical
phenomenon, which has to appear during the fragmentation stage.
However, in the standard implementation of BEC in JETSET, the
fragmentation and decays of the short-lived particles like $\rho$ are
allowed to proceed independently of the Bose-Einstein effect. The BEC
simulation algorithm is applied to the final state particles, for
which the four-momenta difference $Q_{i,j}$ is being calculated for
each pair of identical bosons $i,j$.  A shifted
smaller $Q'_{i,j}$ is then to be found,
such that the ratio $C_2(Q)$ of ``shifted'' to the original $Q$
distribution is given by the requested parameterization (Gaussian or
exponential). In our case, the Gaussian parameterization identical to
the form (\ref{eq:cfpar}) was used~:
\be
\label{eq:c2jetset}
C_2(Q)=1+\lambda e^{-R_{inp}^2Q^2}\ ,
\ee
where $\lambda$ and $R_{inp}$ are input parameters of the model. The input
value of $\lambda$ is often set to 1, as it was done in this
analysis too. Values of $R_{inp}$ usually are chosen to fit
experimental results.

Further, under assumption of a spherical phase space,  $Q'$ is the 
 solution of the equation~:
\be
\label{eq:qprime}
\int\limits_0^{Q_{i,j}}\frac {Q^2 dQ} {\sqrt{Q^2+4m^2}}=\int\limits_0^{Q'_{i,j}}
C_2(Q)\frac {Q^2 dQ} {\sqrt{Q^2+4m^2}}\ .
\ee

After applying corresponding four-momentum shift to each pair of
considered bosons, all particle momenta are re-weighted to satisfy the
energy-momentum conservation. This built-in JETSET algorithm works
only in terms of the invariant four-momenta difference $Q$, i.e., it
does not distinguish between different $Q$ components. This is yet
another ambiguous assumption of the model. Also, it does not include
particle correlations of higher orders.

Evidently, this algorithm is absolutely phenomenological and is not
based on any fundamental theory. It solely changes the final state
particles momenta in order to resemble presence of the Bose-Einstein
correlations. Moreover, presumption of the spherical shape for the
phase space in Eq.(\ref{eq:qprime}) is correct only for the case of very
low $Q$ (see the following discussion).

In spite of all the ambiguities, JETSET reproduces fairly well
experimental data, such as shift of the $\rho$ mass and observed
Bose-Einstein correlations in terms of $Q$. It is widely used to
calculate acceptance corrections for detectors and for various
estimations, like the $W$ mass shift mentioned in the Introduction. To
our mind, this peculiarity is worth investigating, if not in order to
get better understanding of the BEC influence in experimental data,
then at least in order to establish limits of applicability of such a
simulation model.


\section{Analysis and discussion}
\label{sec:analysis}

One of the most puzzling inconsistencies in the JETSET simulation of
BEC is that the input shape of (\ref{eq:c2jetset}) can not be obtained
with the same parameters by fitting the resulting measured $C_2(Q)$
with formula (\ref{eq:cfpar}) (see, for example, article by
Fia{\l}kowski and Wit~\cite{fialkowski}). This is mostly due to the
improper phase space approximation in Eq.(\ref{eq:qprime}). However,
this approximation can still be valid for certain input boson source
size $R_{inp}$. To find out whether it is true, we studied $C_2(Q)$ for
different input values of $R_{inp}$ in formula (\ref{eq:c2jetset}).

\begin{figure}[ht]
\begin{minipage}{0.5\linewidth}
\begin{center}
\vspace*{-0.5cm}
\mbox{\psfig{file=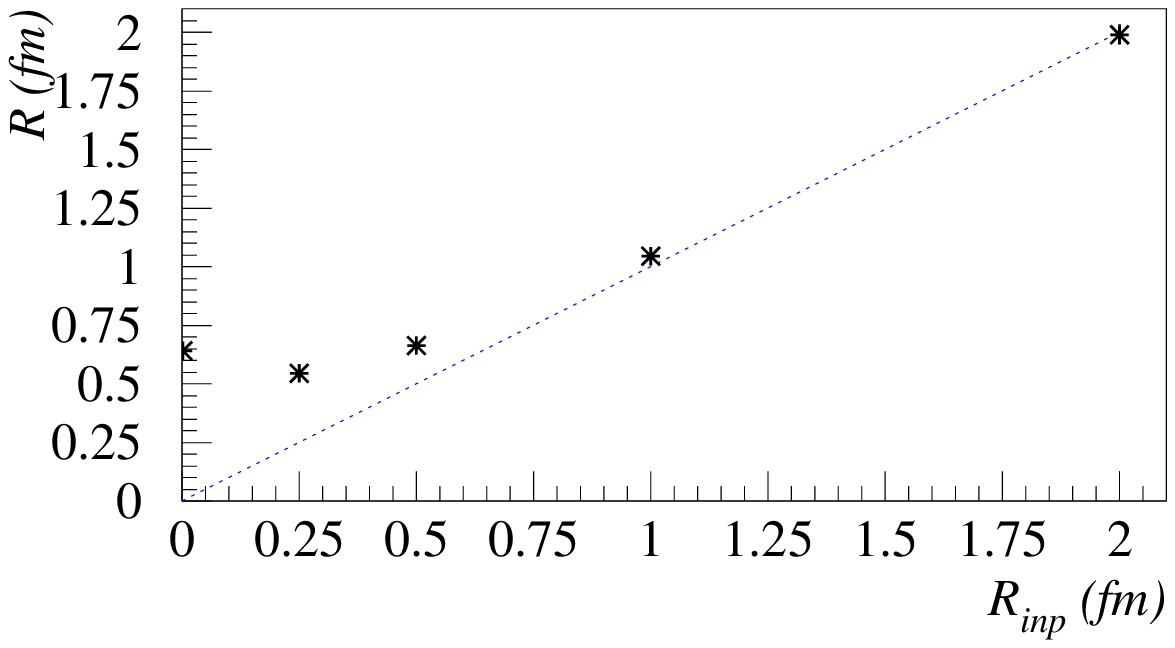,width=\linewidth}}
\vspace*{-1cm}
\caption{\label{fig:rinout1} Width $R$ of the measured in JETSET
  generated events correlation function $C_2(Q)$ as a function
  of the input source radii $R_{inp}$. Line represents the expected dependence.}
\end{center}
\end{minipage}
\begin{minipage}{0.45\linewidth}
\begin{flushright}
{\footnotesize
\begin{tabular}{l|l}
$R_{inp} , fm$ & $R , fm$\\ \hline
0.002 & $0.640 \pm 0.005$\\
0.250 & $0.545 \pm 0.006$\\
0.500 & $0.663 \pm 0.005$\\
1.000 & $1.046 \pm 0.010$\\
2.000 & $1.990 \pm 0.042$\\
\hline DELPHI & $0.489 \pm 0.010$\\
\end{tabular}
}
\end{flushright}
\vspace*{1cm}
\end{minipage}
\end{figure}

Measured as the double-ratio~(\ref{eq:double}) two-particle
correlation function generated with different input source size
$R_{inp}$ was fitted by the form~(\ref{eq:cfpar}). The $\lambda$
parameter always was reproduced at values close to 1, due to the high
purity of the sample. The output source size $R$, however,
behaved differently, see Fig.~\ref{fig:rinout1} and the corresponding
table.  Preliminary DELPHI results measured in 1991-1995 on all the
charged particles are shown in the same table for comparison.
\begin{figure}
\begin{minipage}{0.46\linewidth}
\begin{center}
\vspace*{-0.5cm}
\mbox{\psfig{file=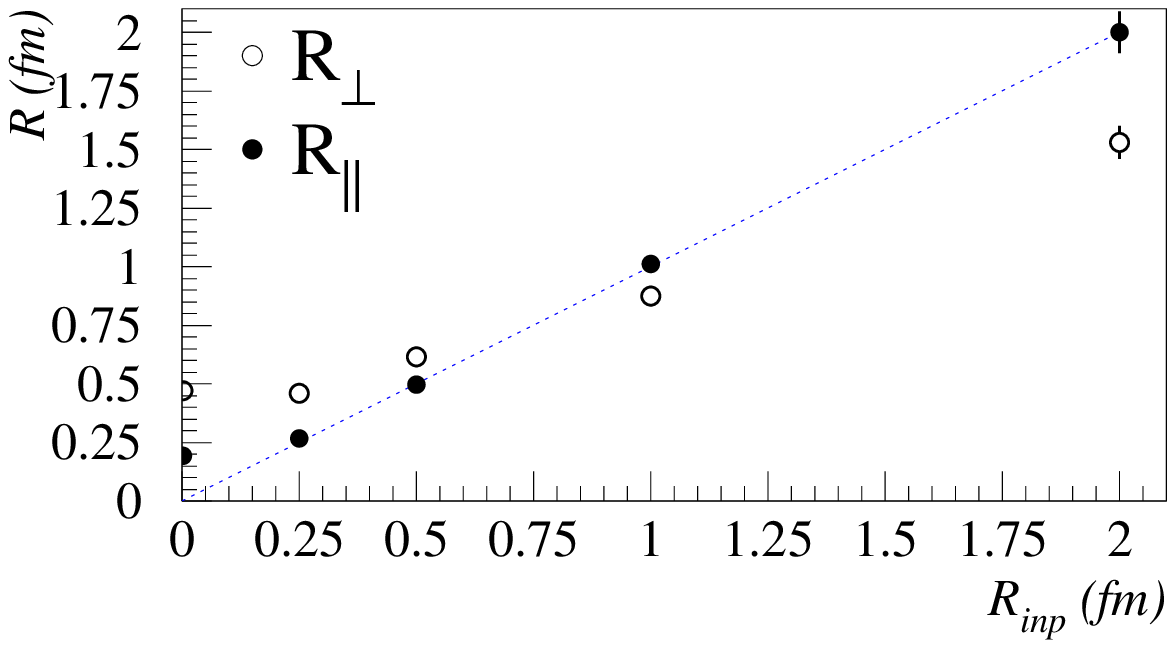,width=\linewidth}}
\vspace*{-1cm}
\\a)\\
\end{center}
\end{minipage}
\begin{minipage}{0.46\linewidth}
\begin{center}
\vspace*{-0.5cm}
\mbox{\psfig{file=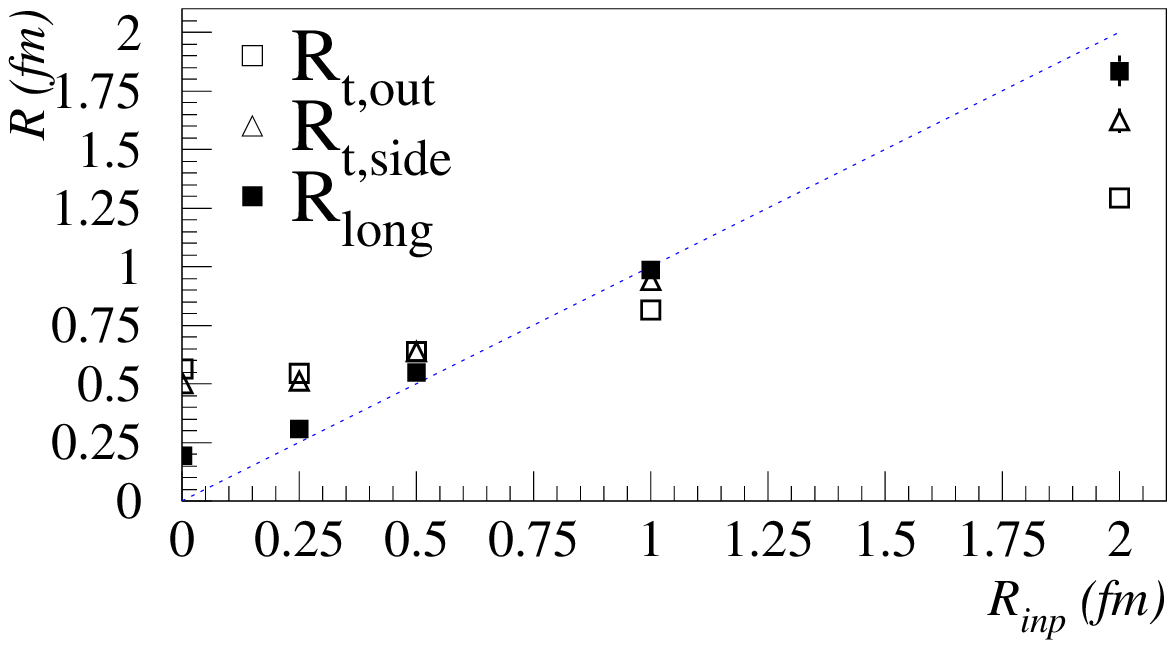,width=\linewidth}}
\vspace*{-1cm}
\\b)\\
\end{center}
\end{minipage}
\caption{\label{fig:rinout23}  Components of the correlation width as a function of the input source radii $R_{inp}$~:
  a) $R_{\parallel}$ , $R_{\perp}$ and b) $R_{t,out}$, $R_{t,side}$,
  $R_{long}$.}
\end{figure}
\begin{flushleft}
{\footnotesize
\vspace*{-1cm}
\begin{tabular}{l|l|l|l|l|l}
$R_{inp} , fm$ & $R_{\perp} , fm$ & $R_{\parallel} , fm$& $R_{t,out} , fm$ & $R_{t,side} , fm$& $R_{long} , fm$\\ \hline
0.002 & $ 0.472\pm 0.002$&$ 0.193\pm 0.001$&$ 0.562\pm 0.005$&$ 0.499\pm 0.004$&$ 0.193\pm 0.002$ \\
0.250 & $ 0.459\pm 0.003$&$ 0.266\pm 0.002$&$ 0.544\pm 0.004$&$ 0.511\pm 0.004$&$ 0.307\pm 0.003$ \\
0.500 & $ 0.616\pm 0.005$&$ 0.497\pm 0.005$&$ 0.637\pm 0.005$&$ 0.636\pm 0.005$&$ 0.549\pm 0.004$ \\
1.000 & $ 0.873\pm 0.015$&$ 1.013\pm 0.021$&$ 0.814\pm 0.011$&$ 0.940\pm 0.012$&$ 0.986\pm 0.013$ \\
2.000 & $ 1.530\pm 0.070$&$ 2.000\pm 0.090$&$ 1.291\pm 0.039$&$ 1.622\pm 0.051$&$ 1.835\pm 0.065$ \\\hline 
DELPHI
      & $ 0.300\pm 0.040$&$ 0.640\pm 0.020$&$ 0.364\pm 0.009$&$ 0.173\pm 0.019$&$ 0.628\pm 0.008$ \\
\end{tabular}
}
\vspace*{0.5cm}
\end{flushleft}

It is clearly seen that the measured $R$ does not depend on the
input $R_{inp}$ when the latter is below $\approx 0.6\,fm$. For the
higher values of $R_{inp}$ JETSET basically reproduces the demanded
correlation function.

Knowing that JETSET does not distinguish between components of
invariant momentum difference $Q$, we should expect similar behaviour
of radius parameters of two- and three-dimensional correlation
functions. Parameterization of these functions is performed in a form
of a multi-dimensional Gaussians (\ref{eq:multi2}) and
(\ref{eq:multi3}) correspondingly.

As one can see from Fig.~\ref{fig:rinout23} and in the corresponding table,
transverse radii follow the same pattern as the $R$, while the
longitudinal radius tends to reproduce the input value of
$R_{inp}$.

\begin{figure}[hb]
\begin{center}
\vspace*{-0.5cm}
\mbox{\psfig{file=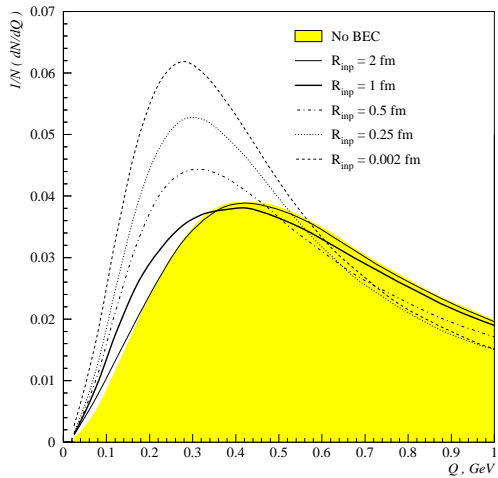,width=0.6\linewidth}}
\vspace*{-0.5cm}
\caption{\label{fig:peaks}  Distribution of $Q$ as a function of the
  input source radii $R_{inp}$. Shaded area shows this distribution in
  absence of BEC.}
\end{center}
\end{figure}

All these results show that there is a certain mechanism in the model,
which imposes lower limit of around $0.6\,fm$ onto measured $R$
and its transverse components, and almost does not affect the
longitudinal radius. The explanation of this phenomenon is illustrated
at Fig.~\ref{fig:peaks} for the one-dimensional case. It shows
evolution of the $dN/dQ$ distribution with input source radius
$R_{inp}$ in comparison with the original non-correlated distribution.
It is clearly seen that the expected Bose-Einstein enhancement appears
only to the left of the non-correlated distribution peak,
$Q<0.3\,GeV$. Since the model conserves multiplicity, and because the
assumption of the spherical phase space in Eq.(\ref{eq:qprime}) is
valid only for this region of the linearly increasing $dN/dQ$, a
depletion appears for $Q>0.3\,GeV$, which results in a non-Gaussian
output correlation function.

\begin{figure}[ht]
\begin{center}
\vspace*{-0.5cm}
\mbox{\psfig{file=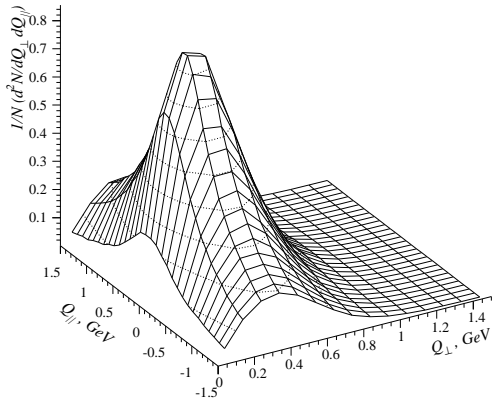,width=0.6\linewidth}}
\vspace*{-0.5cm}
\caption{\label{fig:qtql}  Two-dimensional distribution
  $d^2N/dQ_{\perp}\,dQ_{\parallel}$ in absence of BEC.}
\end{center}
\vspace*{-0.5cm}
\end{figure}

Therefore, the position of the peak in the non-correlated $dN/dQ$
distribution constitutes the limitation of the measured correlation
width $R$ and can be interpreted as a new length scale. This
conclusion is also valid for transverse correlation radii (see
Fig.~\ref{fig:qtql}). In the longitudinal direction, $dN/dQ_{\parallel}$
has less rapid falloff and peaks at a very small $Q_{\parallel}$ value due
to the LCMS properties, and thus is virtually insensitive to the
mentioned length scale.

 As a result, one should state that the built-in JETSET model for
simulating BEC is fully applicable only for sufficiently big sizes of
boson source~: above $1\,fm$. At the same time, experimental data
indicate that this size is around $0.5\,fm$ at the $Z$
peak~\cite{LEP}. This means that, at first, this kind of model
has no predictive power. At second, one should be very careful when
using JETSET with this model for calculation of detector corrections,
because it can not produce an adequate unfolding matrix.

It has to be mentioned that nowadays some other models for simulation
of BEC are being developed~\cite{markus,sharka}. They use direct
implementation of the Bose-Einstein interference into the string
model, being theoretically accurate in this sense. At the moment they
take too much computing resources to be used by high-energy physics
experiments, but they do have predictive power. One of the most
interesting predictions is that the transverse component of the boson
source size, $R_{\perp}$, must be significantly smaller then the
longitudinal one, $R_{\parallel}$. As one can see, it is being
confirmed by the preliminary DELPHI results, but it is not the case
for the present JETSET version. This must encourage further works
towards developing and implementing advanced models for the BEC
simulation in particle generators.


\section*{Acknowledgments}

We would like to thank T.~Sj\"{o}strand for valuable help during this
analysis. Some of us (O.~S. and B.~L.) are grateful to the organisers
of the ``Correlations and Fluctuations 98'' workshop for their
hospitality and for creating an outstanding working environment.



\section*{References}
\begin{thebibliography}{99}

\bibitem{ww1} 
  L.~L\"{o}nnblad and T.~Sj\"{o}strand, \Journal{\PLB}{351}{293}{1995}.

\bibitem{ww2} 
  V.~Kartvelishvili, R.~Kvatadze, R.~M{\o}ller,
  ``Estimating the effects of Bose-Einstein correlations on the $W$
  mass measurement at LEP2'', University of Manchester preprint
  MC-TH-97/04, MAN/HEP/97/1

\bibitem{jetset} 
  T.~Sj\"{o}strand, {\em Comp. Phys. Comm.} {\bf 28}, 229 (1983);\\ 
  T.~Sj\"{o}strand, {\em \textsc{Pythia} 5.6 and \textsc{Jetset} 7.3},
  CERN-TH.6488/92 (1992).

\bibitem{haywood} S.~Haywood, ``Where Are We Going With Bose-Einstein
    -- a Mini Review'', RAL Report RAL-94-074

\bibitem{fialkowski}
  K.~Fia{\l}kowski and R.~Wit, \Journal{\ZPC}{74}{145}{1997}.

\bibitem{bengt} B.~L\"{o}rstad, \Journal{\IJMF}{12}{2861}{1989}.

\bibitem{lcms}
  T.~Cs\"{o}rg\H{o} and S.~Pratt, in ``Proceedings of the Workshop on
  Relativistic Heavy Ion Physics'', KFKI-1991-28/A, p75.

\bibitem{ox2}  B.~L\"{o}rstad, O.~Smirnova: ``Transverse Mass 
Dependence of Bose-Einstein Correlation Radii in $e^+ e^-$ Annihilation at 
LEP Energies'': Proceedings of the 7th International Workshop on 
Multiparticle Production 'Correlations and Fluctuations', June 30 to July 6,
 1996, Nijmegen, The Netherlands.

\bibitem{LEP}
  OPAL Coll.,  P~.D.~Acton et al., \Journal{\PLB}{267}{143}{1991};\\
  DELPHI Coll., P.~Abreu et al., \Journal{\PLB}{286}{201}{1992};\\
  ALEPH Coll.,  D.~Decamp et al., \Journal{\ZPC}{54}{75}{1992}.

\bibitem{markus} 
  B.~Andersson and M.~Ringn\'{e}r, \Journal{\NPB}{513}{627}{1998};\\
  B.~Andersson and M.~Ringn\'{e}r, \Journal{\PLB}{421}{283}{1998}.

\bibitem{sharka} 
  \^{S}.~Todorova-Nov\'a, J.~Rame\u s, `` Simulation of
  Bose-Einstein effect using space-time aspects of Lund string
  fragmentation model'', IReS 97-29, PRA-HEP 97/16.


\end{thebibliography}
\end{document}